\documentclass[12pt]{article}

\usepackage{bm}
\usepackage{amsmath,amssymb,amsthm} 
\usepackage{epsfig}
\usepackage{latexsym} 
\usepackage{multicol}

\newcommand{\lenz}{{\bf L}}
\newcommand{\erre}{{\bf r}}
\newcommand{\erredot}{\dot{\bf r}}

\newcommand{\angmom}{{\bf c}}

\newcommand{\energy}{\mathcal{E}}

\newtheorem{remark}{Remark}

\def\AA{{\bf A}}
\def\BB{{\bf B}}
\def\CC{{\bf C}}
\def\DD{{\bf D}}
\def\EE{{\bf E}}
\def\FF{{\bf F}}
\def\GG{{\bf G}}
\def\JJ{{\bf J}}
\def\cc{{\bf c}}
\def\rhodot{\dot{\rho}}
\def\q{{\bf q}}  
\def\dq{\dot{\bf q}} 

\def\Att{{\cal A}}
\def\alphadot{\dot{\alpha}} 
\def\deltadot{\dot{\delta}} 

\def\br{{\bf r}}   

\def\bPhi{\bm{\Phi}}
\def\bPsi{\bm{\Psi}}

\def\R{\mathbb{R}}

\def\vvec{{\bf v}}
\def\wvec{{\bf w}}

\def\erho{\hat{\bf e}^\rho}
\def\ealpha{\hat{\bf e}^\alpha}
\def\edelta{\hat{\bf e}^\delta}
\def\qrho{\q\cdot\erho}
\def\qalpha{\q\cdot\ealpha}
\def\qdelta{\q\cdot\edelta}
\def\dqrho{\dq\cdot\erho}
\def\dqalpha{\dq\cdot\ealpha}
\def\dqdelta{\dq\cdot\edelta}

\def\diffLenz{\Delta_{\cal L}}

\title{Orbit Determination with the two-body Integrals. II 
}

\author{{\bf Giovanni F. Gronchi,
Davide Farnocchia, Linda Dimare}\\
Dipartimento di Matematica, Universit\`a di Pisa\\
\normalsize{\tt gronchi@dm.unipi.it}\\
\normalsize{\tt farnocchia@mail.dm.unipi.it}\\
\normalsize{\tt dimare@mail.dm.unipi.it}\\
}

\date{}
 
\begin{document}
\maketitle
\begin{abstract}

  The first integrals of the Kepler problem are used to compute preliminary
  orbits starting from two short observed arcs of a celestial body, which may
  be obtained either by optical or radar observations. We write polynomial
  equations for this problem, that we can solve using the powerful tools of
  computational Algebra. An algorithm to decide if the {\em linkage} of two
  short arcs is successful, i.e. if they belong to the same observed body,
  is proposed and tested numerically.
  \noindent In this paper we continue the research started in
  \cite{gronchi&al10}, where the angular momentum and the energy integrals
  were used. A suitable component of the Laplace-Lenz vector in place of the
  energy turns out to be convenient, in fact the degree of the resulting
  system is reduced to less than half.
%

\end{abstract}

\section{Introduction}

We present a new method, based on the first integrals of the Kepler problem,
to compute a finite set of preliminary orbits of a celestial body from two
short arcs of observations.  We assume that the body moves on a Keplerian
orbit with a known center of attraction $O$,\footnote{For asteroid orbits
  $O$ corresponds to the center of the Sun, for space debris $O$ is the center
  of the Earth} and is observed from a point $P$, whose motion is a known
function of time. We deal with two different kinds of observations,
optical and radar, and we make use of the related attributables (see
\cite{ident4}, \cite{tommei&al07}).\footnote{The two different attributables
  can be obtained from observations made from different stations}

In \cite{gronchi&al10} the angular momentum and the energy integrals are used
to solve the linkage problem for solar system bodies. This means to identify
two attributables as related to the same observed object by computing (at
least) one reliable orbit from the observations of both attributables.  The
equations of the problem are written in a polynomial form and the total degree
of the system is 48.  The use of these integrals for the linkage problem has
been first proposed in \cite{taffhall77}, but without fully exploiting the
algebraic character of the problem.

The algorithm presented in \cite{gronchi&al10} has been used in
\cite{farnocchia&al10} for the problem of {\em correlation}\footnote{that is
  the {\em linkage} problem, in the context of space debris} of space debris:
here the authors have extended the method including the oblateness effect
of the Earth.

In this paper we propose different equations for the same problem: in
particular we use a suitable projection of the Laplace-Lenz vector in place of
the energy. The advantage of this approach is that there are several
cancellations and
the total degree is 20.

The same equations can be written using different data, simply considering
other quantities as unknowns: in Section~\ref{s:mixed} we deal with the
case of an optical and a radar attributable. This case is peculiar because we
end up with a univariate polynomial of degree
4. Thus this problem admits {\em explicit} solutions.

In both cases the solutions must fulfill compatibility conditions (as also
shown in~\cite{gronchi&al10}), taking into account the other integrals of
Kepler's problem. To select the solutions we propose a different strategy,
based on the attribution algorithm of a very short arc to a known orbit, see
\cite{ident4}, \cite{milgro09}.  

The structure of the paper is the following.  After introducing some
definitions in Section~\ref{s:att}, we study the linkage of two optical
attributables in Section~\ref{s:2optatt}, while in Section~\ref{s:mixed} we
consider the same problem with one optical and one radar attributable. The
degenerate cases are shown in Section~\ref{s:degen}. Sections~\ref{s:covar}
and \ref{s:select} are devoted to explain the computation of the covariance
matrix for each orbit and the selection of the solutions.  We conclude with a
numerical test in Section~\ref{s:examples}.

\section{Preliminaries}
\label{s:att}

Let us fix an inertial reference frame, with the origin at the center of
attraction $O$.  The position $\q$ and velocity $\dq$ of the observer are
known functions of time.  We describe the position of the observed body as the
vectorial sum
\begin{equation}
\br = \q + \rho\erho\,,
\label{position}
\end{equation}
with $\rho$ the topocentric distance and $\erho$ the {\em line of sight} unit
vector.
We choose spherical coordinates $(\alpha, \delta, \rho)\in [-\pi,\pi)\times
(-\pi/2,\pi/2)\times \R^+$, so that
\[
\erho = (\cos\delta\cos\alpha,\cos\delta\sin\alpha,\sin\delta)\ .
\]
A typical choice for $\alpha,\delta$ is right ascension and declination.
Then we can write the velocity vector
\begin{equation}
\erredot = \dq + \dot{\rho}\erho +
\rho(\alphadot\cos\delta\ealpha + \deltadot\edelta)\,,
\qquad
\rhodot,\alphadot,\deltadot\in\R, \rho\in\R^+\,,
\label{velocity}
\end{equation}
where $\rhodot$, $\rho\alphadot\cos\delta$, $\rho\deltadot$ are the components
of the velocity, relative to the observer in $P$, in the (positively oriented)
orthonormal basis
$\{\erho,\ealpha,\edelta\}$, with
\[
\ealpha =
(\cos\delta)^{-1}\frac{\partial \erho}{\partial\alpha}\,,\hskip 1cm \edelta =
\frac{\partial \erho}{\partial\delta}\ .
\]

We recall the definitions of optical and radar attributables.
From a short arc of optical observations of a moving body $(t_i, \alpha_i,
\delta_i)$ with $i=1\ldots m$, $m\geq 2$, it is possible to compute
an \textbf{optical attributable}
\[
\Att_{opt} = (\alpha, \delta,
\alphadot, \deltadot) \in [-\pi,\pi) \times
(-\pi/2,\pi/2) \times \R^2\,,
\]
representing the angular position and velocity of the body at a mean time
$\bar t$ (see \cite{ident4},\cite{gronchi&al10}).  In this case the radial
distance and velocity $\rho, \rhodot$ are completely undetermined and are
the missing quantities to define an orbit for the body.

\noindent From a set of radar observations of a moving body $(t_i, \alpha_i,
\delta_i, \rho_i)$, with $i=1\ldots m$, $m\geq 2$, it is possible to compute a
\textbf{radar attributable}, i.e. a vector
\[ 
{\Att}_{rad}=(\alpha,\delta,\rho,\dot\rho) \in [-\pi,\pi) \times
(-\pi/2,\pi/2) \times \mathbb R^+ \times \mathbb R\, ,
\]
at time $\bar{t}$ (see \cite{tommei&al07}).  Here $\alphadot,\deltadot$ are
the unknowns needed to define an orbit.

\noindent We call {\bf attributable coordinates} the vector
$(\alpha,\delta,\alphadot,\deltadot,\rho,\rhodot)$ representing the position
and velocity of the body as seen from the observer at time $\bar t$.

\section{Linking two optical attributables}
\label{s:2optatt}

Given two optical attributables $\Att_1, \Att_2$ at epochs $\bar t_1, \bar
t_2$, we assume they belong to the same observed body and write 4 scalar
algebraic equations for the topocentric distances $\rho_1, \rho_2$ and the
radial velocities $\rhodot_1, \rhodot_2$ at the two epochs.

\noindent We use some of the algebraic integrals of the Kepler problem,
i.e. the angular momentum $\angmom$,
and the Laplace-Lenz vector $\lenz$.  The expressions of these integrals
as functions of the topocentric distance and radial velocity $\rho,\rhodot$
are given below.
 
\medbreak\noindent{\sc Angular momentum:}
\[
\cc(\rho,\rhodot) = \br \times \erredot = \DD \rhodot + \EE \rho^2 + \FF \rho
+ \GG\,,
\]
where
\[
\begin{array}{l}
\DD = \q\times\erho\,,\cr
\EE = \alphadot\cos\delta\erho\times\ealpha +
\deltadot\erho\times\edelta 
= \alphadot\cos\delta\edelta - \deltadot\ealpha \,,\cr
\FF = \alphadot\cos\delta\q\times\ealpha + \deltadot\q\times\edelta + \erho\times\dq\,,\cr
\GG = \q\times\dq\ .\cr
\end{array}
\]


\medbreak\noindent{\sc Laplace-Lenz's vector:}
\[
\mu\lenz(\rho,\rhodot) = \erredot\times\angmom -
\mu\frac{\erre}{|\erre|} = \Bigl(\vert\erredot\vert^2
  -\frac{\mu}{|\erre|}\Bigr)\erre - (\erredot\cdot\erre)\erredot\,,
\]
where $\mu$ is a positive constant\footnote{$\mu = Gm_\odot$ if we deal with
  objects orbiting around the Sun; $\mu=Gm_\oplus$ for satellites of the
  Earth.} and
\begin{eqnarray*}
|\erre| &=& (\rho^2 + |\q|^2 + 2\rho\qrho)^{1/2}\,, \\
\vert\erredot\vert^2 &=& \rhodot^2 
+ (\alphadot^2\cos^2\delta + \deltadot^2)\rho^2 + 2\dqrho\rhodot +
2\dq\cdot(\alphadot\cos\delta\ealpha + \deltadot\edelta)\rho + \vert\dq\vert^2
\,,\\
\erredot\cdot\erre &=& \rho\rhodot + \qrho\rhodot + 
(\dqrho + \qalpha\alphadot\cos\delta + \qdelta\deltadot)\rho +  \dq\cdot\q\ .
\end{eqnarray*}

\begin{remark} If $O$ corresponds to the center of the Sun, then we use
  interpolated values for $\q,\dq$, as suggested by Poincar\'e
  \cite{poincare1906}. If $O$ corresponds to the center of the Earth we do not
  apply this method.
\end{remark}

\noindent These dynamical quantities give 6 scalar integrals of the motions:
only 5 are mutually independent, in fact we have
$\lenz\cdot\angmom = 0$.
%
Since we have 4 unknowns, generically we only need 4 scalar conservation laws
to define a finite number of solutions. We select the conservation of the
angular momentum vector and of a particular component of the Laplace-Lenz
vector. The choice of the latter integral presents a substantial advantage
with respect to the use of the energy, as in \cite{gronchi&al10}: the
difference between the two choices will be discussed later.

\subsection{The polynomial equations}
\label{s:polysys}

We use the notation above, with index 1 or 2
referring to the epoch.
If $\Att_1$, $\Att_2$ correspond to the same observed object, 
then the angular momentum vectors at the two epochs must coincide:
\begin{equation}
\angmom_1(\rho_1,\rhodot_1) = \angmom_2(\rho_2,\rhodot_2)\ .
\label{angmomeq}
\end{equation}
Equation (\ref{angmomeq}) can be written as
\begin{equation}
\DD_1\rhodot_1 - \DD_2\rhodot_2 = \JJ(\rho_1,\rho_2)\,,
\label{eq_AM}
\end{equation}
where 
\[
\JJ(\rho_1,\rho_2) = \EE_2\rho_2^2 - \EE_1\rho_1^2 + \FF_2\rho_2 -
\FF_1\rho_1 + \GG_2 - \GG_1\ .
\]
Following \cite{gronchi&al10} we eliminate the variables $\rhodot_1,
\rhodot_2$ and obtain the equation
\begin{equation}
\DD_1\times\DD_2\cdot\JJ(\rho_1,\rho_2) = 0\ .
\label{eq_AM_scal}
\end{equation}
We can write the left-hand side of (\ref{eq_AM_scal}) as 
\begin{equation}
q(\rho_1,\rho_2) \stackrel{\rm def}{=} q_{20}\rho_1^2 + q_{10}\rho_1 +
q_{02}\rho_2^2 + q_{01}\rho_2 + q_{00}\,,
\label{quad_form}
\end{equation}
with
\[
\begin{array}{l}
q_{20} = -\EE_1\cdot\DD_1\times\DD_2\,,\cr
q_{10} = -\FF_1\cdot\DD_1\times\DD_2\,,\cr
\end{array}
\hskip 1cm
\begin{array}{l}
q_{02} = \EE_2\cdot\DD_1\times\DD_2\,,\cr
q_{01} = \FF_2\cdot\DD_1\times\DD_2\,,\cr
\end{array}
\]
\[
q_{00} = (\GG_2-\GG_1)\cdot\DD_1\times\DD_2\ .
\]

\smallbreak
The radial velocities are given by
\begin{equation}
\small
\rhodot_1(\rho_1,\rho_2) =
\frac{(\JJ\times\DD_2)\cdot(\DD_1\times\DD_2)}
{|\DD_1\times\DD_2|^2}\,,
\hskip 0.4cm
\rhodot_2(\rho_1,\rho_2) =
\frac{(\JJ\times\DD_1)\cdot(\DD_1\times\DD_2)}
{|\DD_1\times\DD_2|^2}\ .
\label{rhojdot}
\end{equation}

Also the Laplace-Lenz vectors at the two epochs must coincide.  We equate the
projection of both vectors along $\vvec = \erho_2\times\q_2$:
\begin{equation}
\lenz_1(\rho_1,\rhodot_1)\cdot\vvec =
\lenz_2(\rho_2,\rhodot_2)\cdot\vvec \ .
\label{projlenz}
\end{equation}
Actually the projection of $\lenz_2$ along $\vvec$ is particularly simple:
\[
\mu\lenz_2\cdot\vvec =
-(\erredot_2\cdot\erre_2)(\erredot_2\cdot\vvec)\,,
\]
thus (\ref{projlenz}) becomes
\begin{equation}
  \Bigl(\vert\dot\br_1\vert^2
  -\frac{\mu}{|\br_1|}\Bigr)(\br_1\cdot\vvec) - 
  (\dot\br_1\cdot\br_1)(\dot\br_1\cdot\vvec) = 
  - (\dot\br_2\cdot\br_2)(\dot\br_2\cdot\vvec)\ .
\label{eq_lenz}
\end{equation}
After substituting (\ref{rhojdot}), this is an algebraic equation in
$\rho_1,\rho_2$.
Rearranging the terms in (\ref{eq_lenz}) and squaring we obtain
\begin{equation}
\small
p(\rho_1,\rho_2)\stackrel{def}{=}
\mu^2(\br_1\cdot\vvec)^2 - \vert\br_1\vert^2\left\{
\left[\vert\dot\br_1\vert^2 \br_1 - (\dot\br_1\cdot\br_1)\dot\br_1 +
(\dot\br_2\cdot\br_2)\dot\br_2\right]\cdot\vvec\right\}^2 = 0 \ .
\label{polyp}
\end{equation}
This is a polynomial equation of degree 10 in $\rho_1, \rho_2$: in fact, the
projection
\begin{equation}
\begin{array}{ll}
  \dot\br_2\cdot\vvec &= \q_2\cdot
  (\rho_2(\alphadot_2\cos\delta_2 \ealpha_2 + \deltadot_2\edelta_2) +\dq_2)
  \times\erho_2 =\cr
  &= \rho_2(-\alphadot_2\cos\delta_2\q_2\cdot\edelta_2 -
  \deltadot_2\q_2\cdot\ealpha_2) + \erho_2\cdot\q_2\times\dq_2\cr
\end{array}
\label{rdotv}
\end{equation}
does not depend on $\rhodot_2$ and, in the difference 
$\vert\dot\br_1\vert^2 \br_1 - (\dot\br_1\cdot\br_1)\dot\br_1$,
the second degree term in $\rhodot_1$ (i.e. $\rhodot_1^2\rho_1\erho_1$)
cancels out.

\noindent Therefore, to solve the linkage problem, we can consider the
polynomial system
\begin{equation}
\left\{
\begin{array}{l}
p(\rho_1,\rho_2) = 0\cr
q(\rho_1,\rho_2) = 0\cr
\end{array}
\right.\,,
\hskip 1cm
\rho_1, \rho_2>0\ .
\label{intersec}
\end{equation}
with total degree 20. This shows the advantage of this method compared with
the one in \cite{gronchi&al10}, which gives 
total degree 48.

\subsection{Computation of the solutions}
\label{s:compsol}

To compute the solutions of (\ref{intersec}) we define an algorithm similar to
the one in \cite{gronchi2002}, \cite{gronchi2005}, \cite{gronchi&al10}.
By grouping the monomials with the same power of $\rho_2$ we write
\begin{equation}
p(\rho_1,\rho_2) = \sum_{j=0}^{8} a_j(\rho_1)\;\rho_2^j\,,
\hskip 1cm\mbox{where}
\label{lenzpoly}
\end{equation}
\[
\deg(a_j)=\left\{
\begin{array}{lll}
10 &\mbox{ for }j=0  &\cr
10-(j+1) &\mbox{ for }j=2k-1 &\mbox{ with } k\ge 1\cr
10-j &\mbox{ for }j=2k       &\mbox{ with } k\ge 1\cr
\end{array}
\right.
\]
and
\begin{equation}
q(\rho_1,\rho_2) = b_2\;\rho_2^2 + b_1\;\rho_2 + b_0(\rho_1)
\label{AMpoly}
\end{equation}
for some univariate polynomial coefficients $a_j, b_0$ and constants $b_1,
b_2$.

\noindent We consider the resultant $Res(\rho_1)$ of $p,q$ with respect to
$\rho_2$: it is generically a degree 20 polynomial defined as the determinant
of the $10\times 10$ Sylvester matrix
\begin{equation}
{\tt S}(\rho_1) = 
\left(
\begin{array}{ccccccc}
a_{8} &0      &b_2    &0      &\ldots &\ldots &0      \cr
a_{7} &a_{8} &b_1    &b_2    &0      &\ldots &0      \cr
\vdots &\vdots &b_0    &b_1    &b_2    &\ldots &\vdots \cr
\vdots &\vdots &0      &b_0    &b_1    &\ldots &\vdots \cr
a_0    &a_1    &\vdots &\vdots &\vdots &b_0    &b_1    \cr
0      &a_0    &0      &0      &0      &0      &b_0    \cr
\end{array}
\right)\ .
\end{equation}
The positive real roots of $Res(\rho_1)$ are the only possible values of
$\rho_1$ for a solution $(\rho_1,\rho_2)$ of (\ref{intersec}).
\begin{remark}
  The resultant of $p, q$ with respect to $\rho_1$ leads to compute
  determinants of $12\times 12$ matrices, thus the elimination of $\rho_2$ is
  more convenient. On the other hand, if we project the
  Laplace-Lenz vectors on $\erho_1\times\q_1$, it is better
  to eliminate $\rho_1$.
\end{remark}

\noindent Following \cite{gronchi&al10}, we compute the coefficients of
$Res(\rho_1)$ by an evaluation-interpo\-la\-tion method based on the FFT, and then
the roots $\rho_1(k)$ of $Res(\rho_1)$ by the algorithm described in
\cite{bini97}.  The computation of the preliminary orbits is concluded as
follows:
\begin{itemize}
\item[1)] solve the equation $q(\rho_1(k),\rho_2) = 0$;
\item[2)] discard spurious solutions, that is pairs $(\rho_1,\rho_2)$ solving
  (\ref{polyp}) but not (\ref{eq_lenz});
\item[3)] compute the values of $\rhodot_1(k), \rhodot_2(k)$ by (\ref{rhojdot});
\item[4)] write the corresponding orbital elements. The related epochs are $t_i
  = {\bar t}_i - \rho_i(k)/c$, $i=1,2$ where $c$ is the velocity of light
  (aberration correction).
\end{itemize}

\section{Linking radar and optical attributables}
\label{s:mixed}

Assume we have a \textbf{radar attributable} ${\cal
  A}_{rad}=(\alpha,\delta,\rho,\dot\rho)$ at epoch $\bar{t}$.  We introduce
the variables
\[
\xi = \rho\alphadot\cos\delta\,,
\hskip 1cm
\zeta = \rho\deltadot
\]
so that
\begin{eqnarray*}
\erredot &=& \xi\ealpha + \zeta\edelta + (\rhodot\erho + \dq)\,,\\
\vert\erredot\vert^2 &=& 
\xi^2 + \zeta^2 + 2\dqalpha\xi + 2\dqdelta\zeta + |\rhodot\erho + \dq|^2\,,
\\
\erredot\cdot\erre &=& \qalpha\xi + \qdelta\zeta + (\rhodot\erho + \dq)\cdot\erre\ .
\end{eqnarray*}

\noindent The angular momentum as a function of $\xi,\zeta$ is
\begin{eqnarray}
  \mathbf{c}_{rad}(\xi,\zeta) &=& \mathbf A\xi+\mathbf
  B\zeta+\mathbf C\,,\label{angmom_rad}
\label{c_rad}
\end{eqnarray}
where
\[
\mathbf A = \erre\times\ealpha\,,\qquad
\mathbf B = \erre\times\edelta\,,\qquad 
\mathbf C = \erre\times\dq+
\dot\rho\,\q\times\erho \ .
\]

Suppose we have a radar attributable $\Att_{rad}$ at time $\bar{t}_1$ and an
optical attributable $\Att_{opt}$ at time $\bar{t}_2$. Equating the angular
momentum vectors $\angmom_{rad}$ and $\angmom_{opt}$ at the two epochs we
obtain a polynomial system of 3 equations in the 4 unknowns $\xi_1, \zeta_1,
\rho_2, \rhodot_2$:
\begin{equation}
\AA_1\xi_1+\BB_1\zeta_1+\CC_1 =
\DD_2 \rhodot_2 + \EE_2 \rho_2^2 + \FF_2 \rho_2 + \GG_2\ .
\label{linmixed}
\end{equation}
The system is linear in $\xi_1, \zeta_1, \rhodot_2$. By solving for these
variables we obtain
\begin{equation}
\left\{
\begin{array}{rcl}
\xi_1(\rho_2) &=& 
{\sf X}_2\rho_2^2 + {\sf X}_1\rho_2 + {\sf X}_0\cr
\zeta_1(\rho_2) &=& 
{\sf Z}_2\rho_2^2 + {\sf Z}_1\rho_2 + {\sf Z}_0\cr
\rhodot_2(\rho_2) &=& 
{\sf R}_2\rho_2^2 + {\sf R}_1\rho_2 + {\sf R}_0\cr
\end{array}
\right.\ ,
\label{linsol}
\end{equation}
where
\small
\begin{align*}
&{\sf X}_2 = \gamma\,\EE_2\cdot\BB_1\times\DD_2\,,
&&{\sf X}_1 = \gamma\,\FF_2\cdot\BB_1\times\DD_2\,, 
&&{\sf X}_0 = \gamma\,(\GG_2-\CC_1)\cdot\BB_1\times\DD_2\,,\\
&{\sf Z}_2 = -\gamma\,\EE_2\cdot\AA_1\times\DD_2\,, 
&&{\sf Z}_1 = -\gamma\,\FF_2\cdot\AA_1\times\DD_2\,, 
&&{\sf Z}_0 = -\gamma\,(\GG_2-\CC_1)\cdot\AA_1\times\DD_2\,,\\
&{\sf R}_2 = -\gamma\,\EE_2\cdot\AA_1\times\BB_1\,, 
&&{\sf R}_1 = -\gamma\,\FF_2\cdot\AA_1\times\BB_1\,, 
&&{\sf R}_0 = -\gamma\,(\GG_2-\CC_1)\cdot\AA_1\times\BB_1\,,
\end{align*}
\normalsize
and $\gamma = 1/(\AA_1\cdot\BB_1\times\DD_2)$.

\noindent Equating the expressions of the Laplace-Lenz vectors at the two
epochs, and projecting along $\vvec=\erho_2\times\q_2$, yields
\begin{eqnarray*}
&&\mu[\lenz_{rad}(\xi_1,\zeta_1) -
\lenz_{opt}(\rho_2,\dot\rho_2)]\cdot\vvec = \\
&=& \Bigl[\Bigl(\vert\erredot_1\vert^2 - 
\frac{\mu}{\vert\erre_1\vert}\Bigr)\erre_1
- (\erredot_1\cdot\erre_1)\erredot_1\Bigr]\cdot\vvec + 
(\erredot_2\cdot\erre_2)(\erredot_2\cdot\vvec) = 0\ .
\end{eqnarray*}
The term
\begin{eqnarray*}
  \erredot_2\cdot\vvec &=& 
  \rho_2(-\alphadot_2\cos\delta_2\q_2\cdot\edelta_2 + 
  \deltadot_2\q_2\cdot\ealpha_2) +  \erho_2\cdot\q_2\times\dq_2
\end{eqnarray*}
does not depend on $\rhodot_2$ and is linear in $\rho_2$ (cfr. with
(\ref{rdotv})).  Thus, after substituting $\rhodot_2=\dot\rho_2(\rho_2)$,  $\xi_1=\xi_1(\rho_2), \zeta_1=\zeta_1(\rho_2)$ from
(\ref{linsol}), the terms $(\erredot_2\cdot\erre_2)(\erredot_2\cdot\vvec)$ and
$\bigl[\vert\erredot_1\vert^2\erre_1 -
(\erredot_1\cdot\erre_1)\erredot_1\bigr]\cdot\vvec$ are
polynomials of degree 4 in $\rho_2$.
We obtain a univariate polynomial equation 
with degree 4 in $\rho_2$, which admits explicit solutions.  For each positive
root $\rho_2(k)$ we can compute orbital elements at epochs
$t_1=\bar{t}_1-\rho_1/c, t_2=\bar{t}_2-\rho_2(k)/c$ using (\ref{linsol}).

\section{Degenerate cases}
\label{s:degen}

We list the cases that make the equations of the linkage degenerate.

\vskip 0.3cm
{\sc Optical case}\\
The quadratic form (\ref{quad_form}) is completely degenerate
if
\[
\EE_1\cdot\DD_1\times\DD_2 = \EE_2\cdot\DD_1\times\DD_2 = 0\ .
\]
For a discussion on the geometric meaning of these
conditions see \cite{gronchi&al10}.
Another degenerate case occurs if $\erho_2\times\q_2 = 0$. In the case of
space debris this corresponds to a zenith observation.


\vskip 0.3cm
{\sc Radar-Optical case}\\
System (\ref{linmixed}) degenerates if
\[
\AA_1\times\BB_1\cdot\DD_2 = r_1^\rho(\br_1\cdot\DD_2) = 0\ .
\]
This occurs when $\br_1\cdot\erho_1 = 0$, or $\br_1\times \br_2 = {\bf 0}$, or
when $\erho_2$ is in the orbital plane (orthogonal to $\br_1\times\br_2$).
Another degeneration occurs when $\erho_2\times\q_2 = 0$, as in the optical
case.

\section{Covariance of the solutions}
\label{s:covar}

Let $\AA = (\Att_1,\Att_2)$ be the vector of two optical attributables and
$\Gamma_{\AA}$ its covariance matrix.  For each solution ${\bf Y} =
(\rho_1,\rhodot_1,\rho_2,\rhodot_2)$ of the linkage problem
\begin{equation}
\left\{
\begin{array}{l}
  \angmom_1(\rho_1,\rhodot_1) = \angmom_2(\rho_2,\rhodot_2)\cr
\lenz_1(\rho_1,\rhodot_1)\cdot\vvec = \lenz_2(\rho_2,\rhodot_2)\cdot\vvec\cr 
\end{array}
\right.
\hskip 1cm
\rho_1,\rho_2>0\,,
\label{linksys}
\end{equation}
we can compute the Cartesian
coordinates ${\cal E}_{car}^{(1)}$, ${\cal E}_{car}^{(2)}$ at epochs $t_1,
t_2$, and their covariance matrices $\Gamma_{car}^{(1)}$,
$\Gamma_{car}^{(2)}$.
We introduce the following notation:
\begin{itemize}

\item[1)] $\mathbf{E}_{car} =
(\mathcal{E}_{car}^{(1)},\mathcal{E}_{car}^{(2)})$ is the 2 epochs Cartesian
coordinates vector;

\item[2)] $\mathbf{E}_{att} = (\mathcal{E}_{att}^{(1)},
  \mathcal{E}_{att}^{(2)})$, where\footnote{If we use interpolated values for
    $\q,\dot\q$, as suggested in~\cite{poincare1906}, then $\mathcal{E}_{att}^{(i)}$ are not the attributable coordinates corresponding to
    $\mathcal{E}_{car}^{(i)}$, $i=1,2$.}
\[
\mathcal{E}_{att}^{(i)} =
(\alpha_i,\delta_i,\alphadot_i,\deltadot_i,\rho_i,\rhodot_i)\,,\quad i=1,2\ .
\]

\end{itemize}

Define the map $\bPsi: \R^{12} \to \R^4$ by
\[
\mathbf{E}_{car} \stackrel{\bPsi}{\mapsto}  
 \left[
\begin{array}{c}
\angmom_1 - \angmom_2\cr
\mu(\lenz_1-\lenz_2)\cdot \wvec
\end{array}
\right]\ ,
\hskip 1cm
\wvec = \erre_2\times\q_2 \ .
\]

\noindent Moreover, define $\mathcal{T}_{att}^{car}: \mathbf{E}_{att} \to
\mathbf{E}_{car}$ by (\ref{position}), (\ref{velocity}) for both epochs, and
consider the map $\bPhi=\bPsi\circ\mathcal{T}_{att}^{car}$. Then 
$\bPhi=\mathbf{0}$ is equivalent to (\ref{linksys}).\footnote{We use $\wvec$
  instead of $\vvec$ to obtain simpler expressions for the derivatives of
  $\bPhi$.}

%

\noindent The covariance matrix of the Cartesian coordinates at epoch $t_1$ is
\[
\Gamma_{car}^{(1)} = \frac{\partial {\cal E}_{car}^{(1)}}{\partial {\AA}}
\Gamma_{\AA} \left[\frac{\partial {\cal E}_{car}^{(1)}}{\partial
    {\AA}}\right]^T\,,
\]
with
\[
\frac{\partial {\cal
    E}_{car}^{(1)}}{\partial {\AA}} = \frac{\partial {\cal
    E}_{car}^{(1)}}{\partial {\cal E}_{att}^{(1)}} \frac{\partial {\cal
    E}_{att}^{(1)}}{\partial {\AA}}\,, \hskip 1cm \frac{\partial {\cal
    E}_{att}^{(1)}}{\partial {\AA}} = {\small \left[
\begin{array}{c}
\begin{array}{cc}
I_4 &O_4\cr
\end{array}\cr
\displaystyle\frac{\partial(\rho_1, \rhodot_1)}{\partial{\AA}}\cr
\end{array}
\right]}\ .
\]
From the implicit function theorem
\[
\frac{\partial {\bf Y}}{\partial {\AA}}(\AA) =
-\left[\frac{\partial\bPhi}{\partial{\bf
      Y}}(\mathbf{E}_{att})\right]^{-1}\frac{\partial\bPhi}{\partial
  {\AA}}(\mathbf{E}_{att})\,,
\]
where
\[
\frac{\partial\bPhi}{\partial{\bf Y}} = 
\left(\frac{\partial\bPsi}{\partial\mathbf{E}_{car}} 
\circ
{\cal T}_{att}^{car}\right) 
\frac{\partial{\cal T}_{att}^{car}}{\partial{\bf Y}}\,,
\hskip 1cm
\frac{\partial\bPhi}{\partial{\AA}} = 
\left(\frac{\partial\bPsi}{\partial\mathbf{E}_{car}} 
\circ
{\cal T}_{att}^{car}\right) 
\frac{\partial{\cal T}_{att}^{car}}{\partial{\AA}}\ .
\]
The matrices $\frac{\partial{\cal T}_{att}^{car}}{\partial{\bf Y}}$ and
$\frac{\partial{\cal T}_{att}^{car}}{\partial{\AA}}$ are respectively made by
columns 5,6,11,12 and by columns 1,2,3,4,7,8,9,10 of
$\frac{\partial\mathbf{E}_{car}}{\partial\mathbf{E}_{att}}$.

\noindent For a given vector ${\bf u}\in\R^3$ define the
\textbf{hat map}
\[
\R^3\ni (u_1,u_2,u_3) = {\bf u} \mapsto \widehat{\bf u} \stackrel{def}{=}
\left[
\begin{array}{ccc}
0 &-u_3 &u_2\cr
u_3 &0 &-u_1\cr
-u_2 &u_1 &0\cr
\end{array}
\right] \in so(3)\ .
\]
Then we have, using $\hat{\bf u}^T = -\hat{\bf u}$,
\[
\frac{\partial\bPsi}{\partial\mathbf{E}_{car}} = 
\left[
\begin{array}{cccc}
-\widehat{{\erredot}_1}  &\widehat{{\erre}_1}  &\widehat{{\erredot}_2}  &-\widehat{{\erre}_2}\cr
\stackrel{}{\displaystyle\frac{\partial \diffLenz}{\partial \erre_1}} &\displaystyle\frac{\partial \diffLenz}{\partial \erredot_1} &\displaystyle\frac{\partial \diffLenz}{\partial \erre_2} &\displaystyle\frac{\partial \diffLenz}{\partial \erredot_2}\cr
\end{array}
\right]\,,
\]
where

\small
\begin{align*}
&\frac{\partial \diffLenz}{\partial \erre_1}=
\left(|\erredot_1|^2 -
\mu \frac{1}{|\erre_1|}\right)\wvec^T  +
  \mu\frac{(\erre_1\cdot\wvec)}{|\erre_1|^3}\erre_1^T 
- (\erredot_1\cdot\wvec)\erredot_1^T
\,, \\
&\frac{\partial \diffLenz}{\partial \erredot_1} =
2(\erre_1\cdot\wvec)\erredot_1^T - (\erredot_1\cdot\wvec)\erre_1^T -
(\erredot_1\cdot\erre_1)\wvec^T\,, \\
&\frac{\partial \diffLenz}{\partial \erre_2} = \left(|\erredot_1|^2 - \frac{\mu}{|\erre_1|}
\right)[\q_2\times\erre_1]^T - (\erredot_1\cdot\erre_1)[\q_2\times\erredot_1]^T + 
(\erredot_2\cdot\wvec)\erredot_2^T + (\erredot_2\cdot\erre_2)[\q_2\times\erredot_2]^T\,,
\\
&\frac{\partial \diffLenz}{\partial \erredot_2} =
(\erredot_2\cdot\wvec)\erre_2^T + (\erredot_2\cdot\erre_2)\wvec^T\ .
\\
\end{align*}
\normalsize


In the case of one radar and one optical attributable the covariance
of the solutions can be computed in a similar way, with the following
differences:
\begin{itemize}

\item[1)] 
the vector of the attributables is $\AA = (\Att_{rad},\Att_{opt})$,

\item[2)] the vector of unknowns is ${\bf Y} =
  (\alphadot_1,\deltadot_1,\rho_2,\rhodot_2)$,

\item[3)] the matrix of the derivatives of ${\cal E}_{att}^{(1)}$ with respect
  to $\AA$ is
\[
\small\frac{\partial {\cal E}_{att}^{(1)}}{\partial {\AA}} = 
\left[
\begin{array}{c}
\begin{array}{cccc}
 I_2 &O_2 &O_2 &O_2\cr
\end{array}\cr
\displaystyle\frac{\partial(\alphadot_1, \deltadot_1)}{\partial{\AA}}\cr
\begin{array}{cccc}
O_2 &I_2 &O_2 &O_2\cr
\end{array}\cr
\end{array}
\right]\ ,
\]

\item[4)] 
The matrices $\frac{\partial{\cal T}_{att}^{car}}{\partial{\bf Y}}$ and
$\frac{\partial{\cal T}_{att}^{car}}{\partial{\AA}}$ are respectively made by
columns 3,4,11,12 and by columns 1,2,5,6,7,8,9,10 of
$\frac{\partial\mathbf{E}_{car}}{\partial\mathbf{E}_{att}}$.

\end{itemize}

\section{Selecting solutions}
\label{s:select}

The solutions of (\ref{linksys}) are defined by using only four conservation
laws. Thus ${\cal E}_{car}^{(1)}$, ${\cal E}_{car}^{(2)}$ may not correspond
to the same orbit. We select the solutions of the linkage problem by means of
the {\bf attribution} algorithm \cite{ident4}, \cite{milgro09}.
Here we recall briefly the
procedure.

Let ${\cal E}_1$ be a set of orbital elements
for the observed body at time $t_1$, with $6\times 6$ covariance matrix
$\Gamma_1$.  We can propagate the orbit with covariance to the epoch $t_2$ of
an attributable $\Att_2$, with a given $4\times 4$ covariance matrix
$\Gamma_{\Att_2}$, by the formula
\[
\Gamma_2 = \frac{\partial \Phi({\cal E}_1,{\bar
    t}_2)}{\partial {\cal E}_1} \Gamma_1\left[\frac{\partial
    \Phi({\cal E}_1,{\bar t}_2)}{\partial {\cal E}_1} \right]^T
\]
where $\Phi({\cal E}_1,t)$ is the integral flow of the Kepler problem.  Then
we can extract a predicted attributable $\Att_p$, at time $\bar{t}_2$, with
covariance matrix $\Gamma_{\Att_p}$.
Let $C_{\Att_p} = (\Gamma_{\Att_p})^{-1}$ and $C_{\Att_2} = (\Gamma_{\Att_2})^{-1}$.
We define
\[
C_0 = C_{\Att_p} + C_{\Att_2}\,,
\hskip 1cm
\Gamma_0 = C_0^{-1}
\]
The \textbf{identification penalty}
is given by
\[
\chi_4 = (\Att_2 - \Att_p)\cdot[C_{\Att_p} - C_{\Att_p}\Gamma_0
C_{\Att_p}](\Att_2 - \Att_p)\ .
\]
If the value of $\chi_4$ is within a fixed threshold, we
can accept the orbit ${\cal E}_1$.

\begin{remark} To select solutions we could also use compatibility conditions,
  as in \cite{gronchi&al10}.  In this case the conditions could be
\begin{equation}
(\lenz_1-\lenz_2)\cdot\erho_2 = 0\,,
\hskip 1cm
\ell_1 - \ell_2 = n_1(t_1-t_2)\ .
\label{comp_cond}
\end{equation}
\end{remark}

\section{A test case}
\label{s:examples}


We present the results of a test of the method explained in
Section~\ref{s:2optatt} with the asteroid (99942) Apophis.  We take two sets
of 13 and 12 observations respectively with mean epochs 
$\bar{t}_1 = 53175.59$, 
$\bar{t}_2 = 53357.45$.
%
After removing duplicate and spurious solutions
we obtain
\begin{table}[ht]
\begin{center}
\begin{tabular}{l|c|c}
       &$\rho_1$         &$\rho_2$\\
\hline
 1     &0.78987     &0.04345 \\
 2     &1.13777     &0.09569 \\
 \end{tabular}
\end{center}
\caption{Solutions of the system (\ref{intersec}) for ($(99942)$).}
\label{tab:sol_found}
\end{table}

\begin{figure}[!h]
\centerline{\epsfig{figure=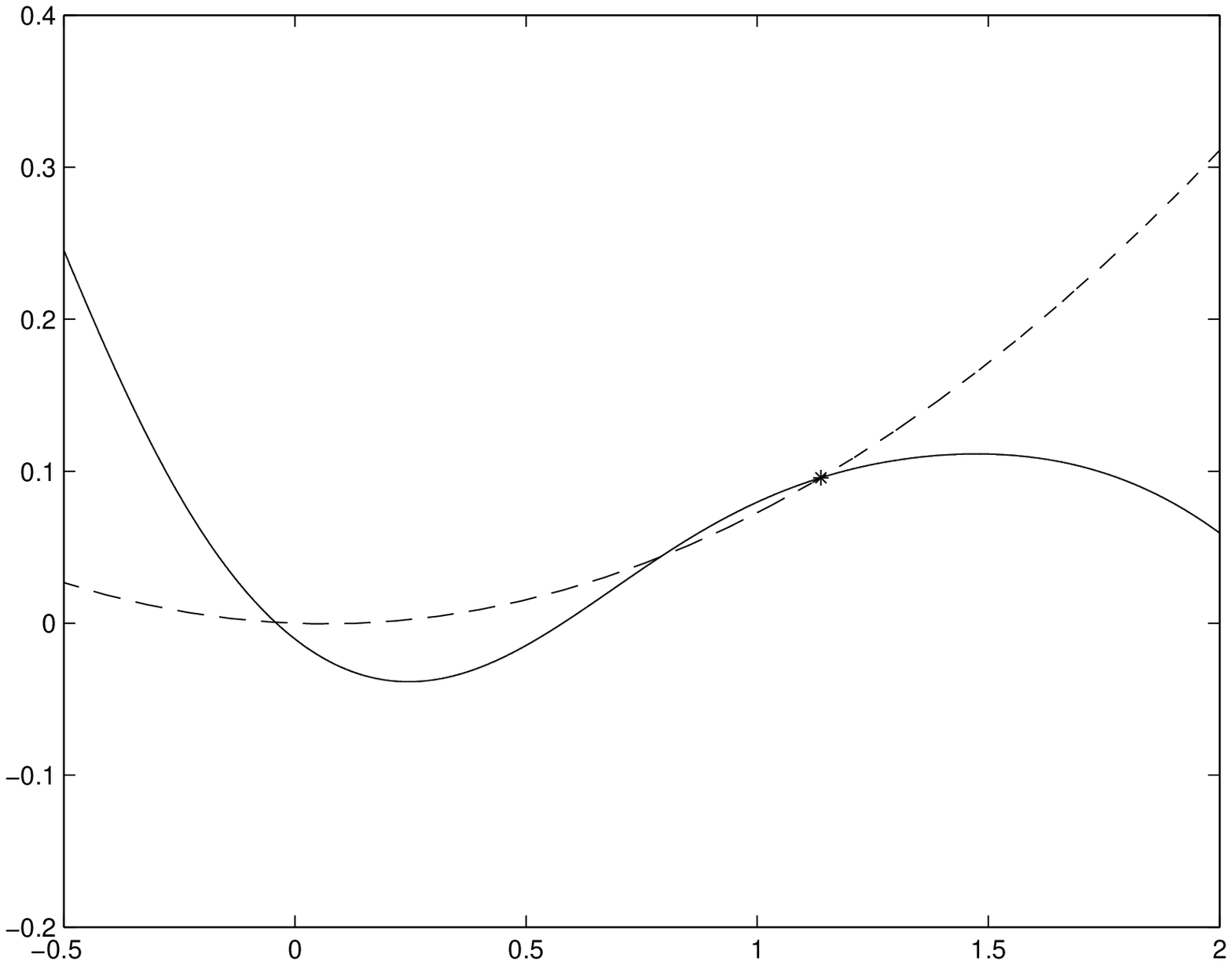,width=6.5cm}
\hskip 0.8cm\epsfig{figure=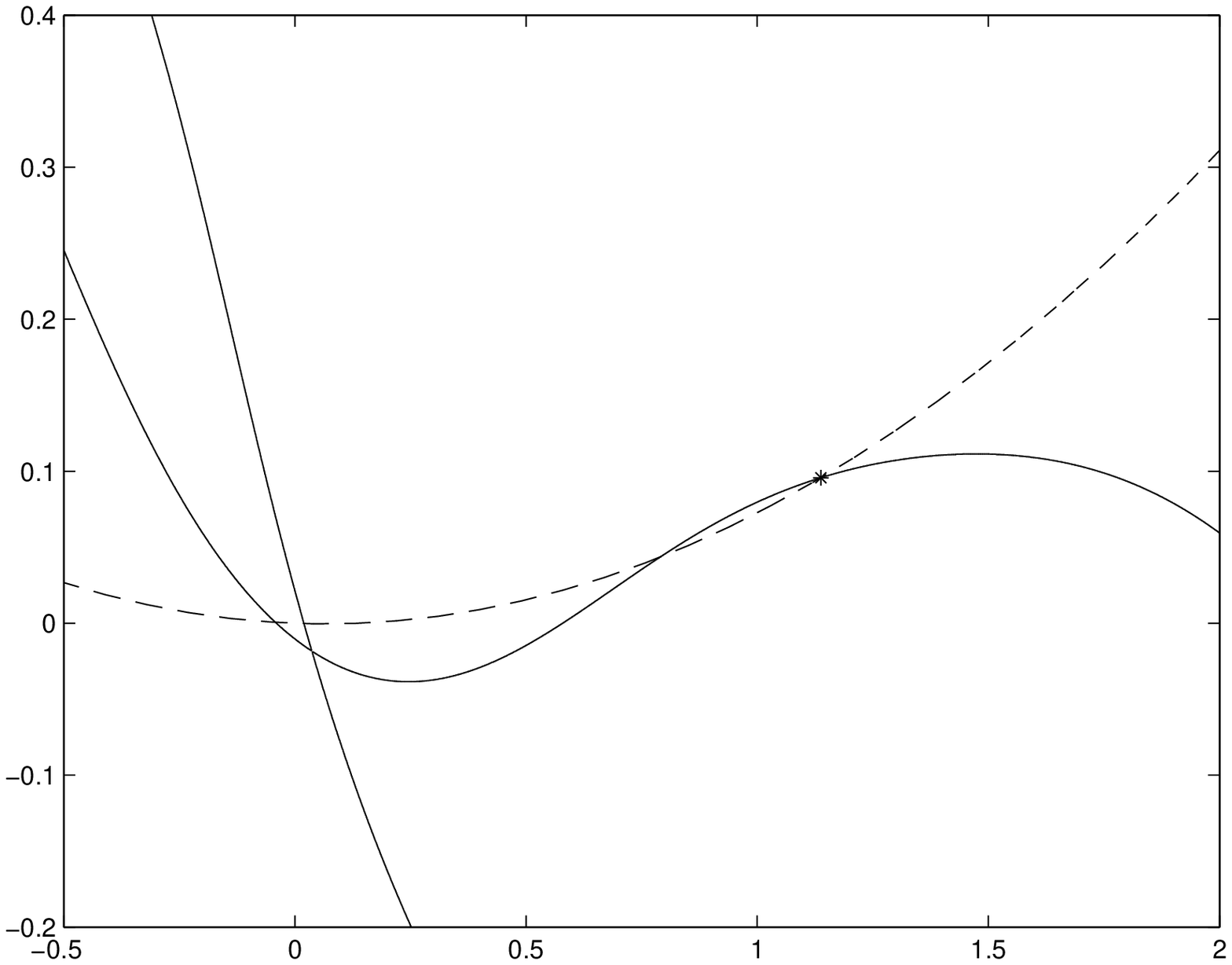,width=6.5cm}}
\centerline{\epsfig{figure=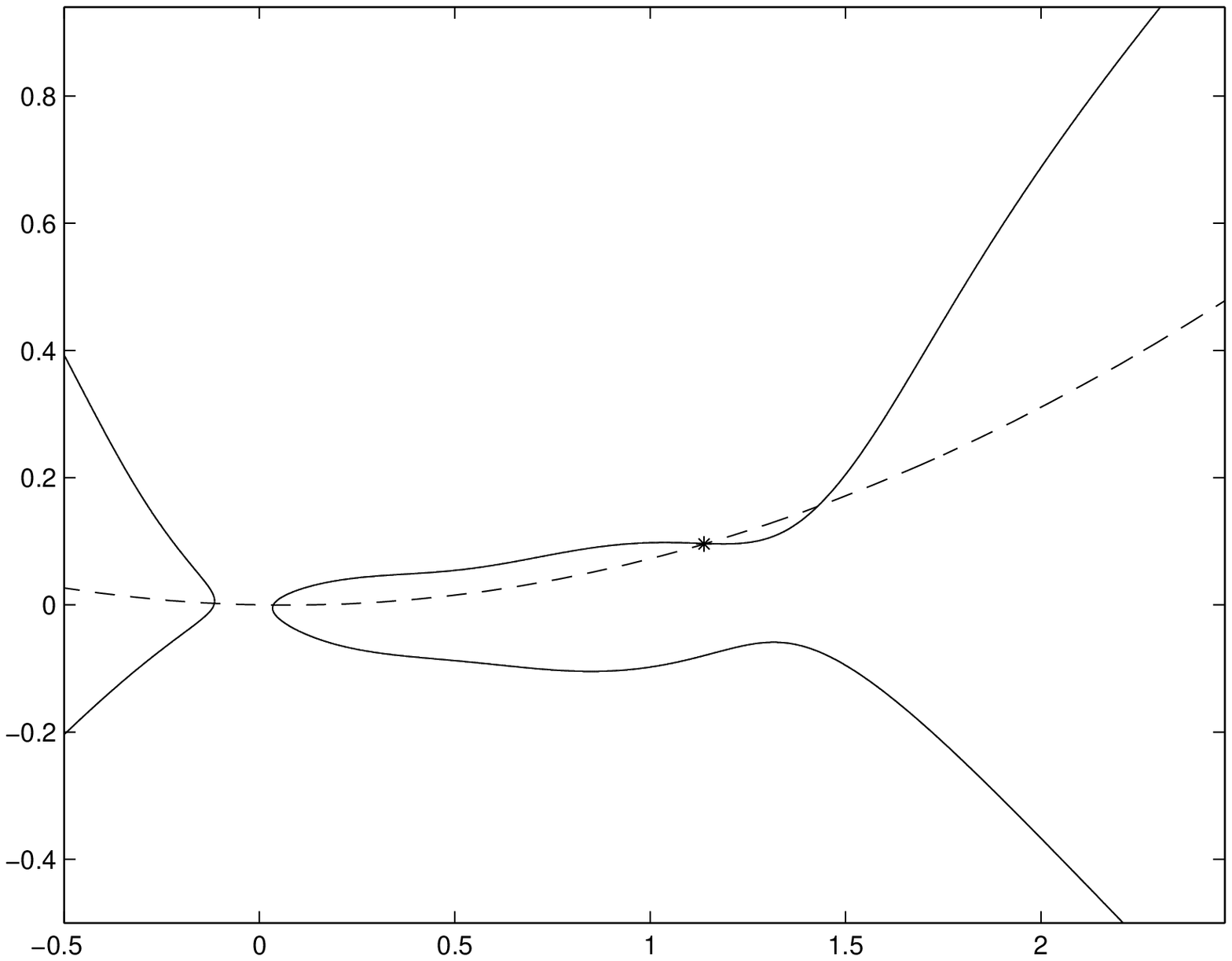,width=6.5cm}
\hskip 0.8cm\epsfig{figure=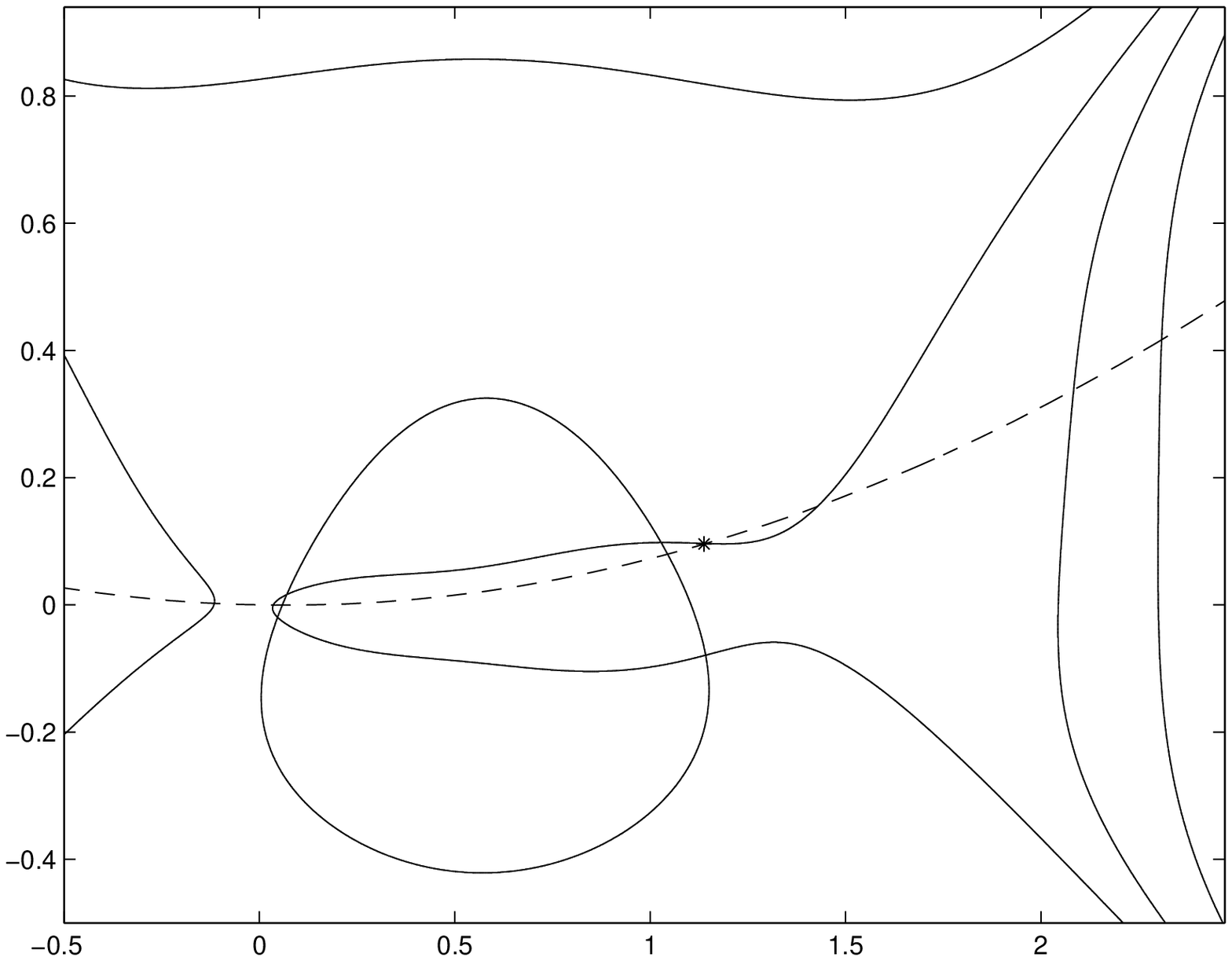,width=6.5cm}}
\caption{\small For the test case of (99942) Apophis, this figure shows the advantage
  of using equation (\ref{projlenz}) instead of the conservation of the energy
  $\energy$. Top left: $\angmom$, $\lenz\cdot\vvec$ integrals.  Top right:
  $\angmom$, $\lenz\cdot\vvec$ integrals, polynomial form. Bottom left:
  $\angmom$, $\energy$ integrals. Bottom right: $\angmom$, $\energy$
  integrals, polynomial form.}
\label{fig:comparison}
\end{figure}

\noindent The two solutions gives respectively $\chi_4(1)= 3230925.94$,
$\chi_4(2)= 2.29$, therefore we select the second one, with Keplerian elements
(distances in AU, angles in degrees)
\[
a = 0.9230\,,
\hskip 0.1cm  e = 0.189\,,  
\hskip 0.1cm  I = 3.287\,,
\hskip 0.1cm  \Omega = 204.912\,,
\hskip 0.1cm  \omega = 124.778\,,
\hskip 0.1cm  \ell =  249.003
\]
at epoch $t_1= 53175.59$.
%
\noindent We can compare the results with the known orbit propagated at epoch
$t_1$:
\[
a = 0.9219\,,        
\hskip 0.1cm e = 0.191 \,,    
\hskip 0.1cm  I =3.333 \,,      
\hskip 0.1cm  \Omega = 204.575 \,,
\hskip 0.1cm  \omega = 126.176 \,, 
\hskip 0.1cm  \ell = 247.500  
\]
%
%
In Figure~\ref{fig:comparison}, for the test case of (99942) Apophis, we show
the intersections of the curves defined in this paper compared with the ones
obtained by the conservation of the energy. In the
four pictures the hyphened curve corresponds to equation
(\ref{eq_AM_scal}). We also draw the curve defined by (\ref{eq_lenz}) on top
left, and the one by (\ref{polyp}), in polynomial form, on top right. The
conservation of the energy defines the curve drawn on bottom left, its
polynomial form (obtained by rearranging terms and squaring twice) defines the
one on bottom right. The orbit determination method introduced in this
paper, searching for the intersections shown on top right, is clearly
convenient with respect to the method investigated in \cite{gronchi&al10},
related to figure on bottom right.




\section{Acknowledgments}
We wish to thank A. Milani for his useful suggestions during the
development of this work.


\end{document}